\documentstyle[12pt,fleqn]{article}
\newcommand{\beq}{\begin{equation}}
\newcommand{\eeq}{\end{equation}}
\newcommand{\beqa}{\begin{eqnarray}}
\newcommand{\eeqa}{\end{eqnarray}}

\newcommand{\ti}{\tilde}

\def\half{\frac{1}{2}}

\def\hs1{\hskip1mm} \def\h10{\hskip10mm} \def\hx{\h10\hbox}
 
\def\<{\langle} \def\>{\rangle}   
  
\def\half{{\scriptstyle{1\over 2}}}

\rm
\begin{document}


\vskip 1cm
\begin{center}
{\large{\bf Generalized Quantum Measurement}} 
\vskip 2cm {Ting Yu and Ian\, C. Percival} \vskip 0.4cm {\sl Department of
Physics, Queen Mary and
Westfield College\\ University of London, Mile End Road, London E1
4NS, UK\\ } \vskip 0.7cm
\end{center}

\vskip 2.4cm

        \date{\today}

\begin{abstract}
We overcome one of Bell's objections to `quantum measurement' by
generalizing the definition to include systems outside the
laboratory. According to this definition a {\sl generalized quantum
measurement} takes place when the value of a classical variable is
influenced significantly by an earlier state of a quantum system.  A
generalized quantum measurement can then take place in equilibrium
systems, provided the classical motion is chaotic. This paper deals
with this classical aspect of quantum measurement, assuming that the
Heisenberg cut between the quantum dynamics and the classical dynamics
is made at a very small scale.  For simplicity, a gas with collisions
is modelled by an `Arnold gas'.

\vskip 0.7cm

PASC number(s): 03.75. Ta; 42.50. Lc

\vskip 0.7cm

   \end{abstract}

\newpage

\section{Introduction}
In one of his last articles \cite{JB}, John Bell made three
charges against quantum `measurement'.  The third of these was: ``
In the beginning natural philosophers tried to understand the
world around them.\dots Experimental science was born.  But
experiment is a tool.  The aim remains: to understand the world.
To restrict quantum mechanics to be exclusively about piddling
laboratory operations is to betray the great enterprise.  A
serious formulation [of quantum mechanics] will not exclude the
big world outside the laboratory.''

A possible answer to this charge is to extend the definition of
quantum measurement into that big world, with particular emphasis
on equilibrium systems \cite{ICPa}.  Traditionally quantum
measurements take place in the laboratory, but the laboratory is
only part of our universe, and all such measurements start out as
imitations of natural phenomena.  Cloud chambers were based on the
physics of clouds, which are natural detectors of charged
particles.  Spark chambers imitate lightning.  We define
{\sl generalized quantum measurement} to mean any process whereby the
state
of a quantum system influences the value of a classical variable
\cite{ICP,ICP1}.  This definition then applies to the big world.

We take the view of those experimenters in the laboratory who never
have any doubt that their apparatus is classical or that quantum
mechanics must be used for the internal dynamics of an atom.  This is
equivalent to assuming that the `Heisenberg cut' between the quantum
and the classical occurs between the two, on small scales.  The
dynamical questions are then moved to the classical domain: whether
small changes like the motion of the atom can affect the behaviour of
large scale classical variables.  This picture, which we apply to
processes outside the laboratory, is consistent with the usual
treatment of amplifiers and detectors in the laboratory, where quantum
mechanics is only used for the early stages.

In this paper, we restrict our attention to Bell's charge
against `quantum measurement' from this viewpoint.  We are not
concerned with other fundamental issues of the quantum measurement
problem(for instance, see \cite{OMM,ICP3} and references therein).

Laboratory quantum measurements include particle states producing
the droplets in cloud chambers, bubbles in bubble chambers and
sparks in spark chambers \cite{WAN}.  They include photon states
producing silver grains in photographic emulsions, and also
electrons and photons producing electron avalanches in solid state
detectors and photomultipliers.

Generalized quantum measurements include ions producing water droplets
in clouds, photon states sending impulses through the optic nerves of
owls and the states of cosmic rays that produced small but very
long-lived dislocations in mineral crystals in the Jurassic era.  This
takeover of the physics of laboratory quantum measurement into the
world outside the laboratory is here generalized, and one of the
questions we have to ask is how far this generalization can go.  Where
else do we find generalized quantum measurements according to our
definition?  In particular, are there generalized quantum measurements
in equilibrium systems?

This problem cannot be solved without a better understanding of
the classical theory of equilibrium systems, in particular the
influence of motion at the atomic scale on variables that are
normally considered to be classical, like sound waves at audio
frequencies, represented by Fourier components of the density of a
gas, which is represented here by a model `Arnold gas'.

The paper is organized as follows. In Section 2, the ideas
underlying our classical model are given.  Subsequently, in
Section 3, we introduce the {\sl Arnold gas} model. It is shown
that this model can be solved analytically. In Section 4, we
present a detailed analysis of the changes in the Fourier
components of the particle density of the molecules in phase space
as a result of collisions which are crucial for a quantitative
description of fluctuation of the gas density.  We conclude in
Section 5 that there is a sense in which there are quantum
measurements in equilibrium gases.

\section{ Equilibrium gases}

Laboratory systems used for quantum measurement are very
complicated physical systems, even stripped down to their bare
essentials. They involve amplification in one form or another, and
so do the natural systems that they imitate.

A gas in equilibrium is simpler, yet we give an example to show that
generalized quantum measurement as we have defined it can take place
there also.  The reason is that the motion of the molecules in the gas
is chaotic, and small changes now result in large changes later.  In
particular changes at the quantum level now produce significant
classical fluctuations in the density later.  However, unlike earlier
examples, we cannot use the classical density fluctuations to learn
anything specific about these earlier quantum states, because the
chaos causes mixing \cite{OTT}, which effectively obscures the signal,
because the initial conditions of other atoms also change the density.
So generalized quantum measurement applied to equilibrium systems does
not have {\sl all} the properties of a laboratory measurement,

In the nineteenth century, Rayleigh recognized that these classical
density fluctuations would scatter light, and that the scattering was
strongly dependent on the wavelength of the light.  The result is the
blue of the sky.  The growth of droplets of water around the charged
particles produced by cosmic rays in the atmosphere is a generalized
quantum measurement.  According to the theory of this paper, so are
the density fluctuations in the atmosphere that cause the sky to be
blue where there are no clouds. So if you ever look at the sky, as
every physicist sometimes should, whether it is clear or overcast, you
are seeing one example or another of generalized quantum measurement.

\section{A soluble model}

In order to understand generalized quantum measurement outside the
laboratory, it is useful to make a detailed analysis of a model.
We consider a classical one-dimensional {\sl Arnold Gas} which can
be analytically solved. In this model, the interaction between two
molecules is represented by the Arnold cat map. We are interested
in the change of Fourier components of probability density at time
$t=T$ due to the initial changes of the state of gas at earlier
time $t=0$. We show that for our model a small change in the state
of a single particle produces a significant density fluctuation in
the gas after a finite time.


\subsection{Collisions and subsystems}

Our model represents a gas of molecules. We want to find the
change in the state of the gas at a time $t=T$ due to an earlier
change in the coordinate and momentum of a single particle $P_0$
at time $t=0$. In order to get a solvable model, some
simplifications and idealizations must be made. To be specific, we
assume the process by which this particle $P_0$ affects the other
particles in stages, without at first considering the time at
which the collisions take place.  The first stage in this process
is the first collision of particle $P_0$ with one other particle
$P_1$, after which this pair of particles are both affected by the
initial coordinate and momentum (state) of the particle $P_0$.
The subsystem $S_1$ after this first stage consists of both
particles of the pair.

In the second stage of the process, each particle of $S_1$
collides with another particle, assumed to be different, giving
the four particles of subsystem $S_2$ affected by the initial
state of $P_0$. Notice that the two collisions of the second stage
need not occur at the same time: questions of timing are
considered later.

Every particle of subsystem $S_n$ belongs to all later subsystems.
We also assume for simplicity that every particle of $S_n$
collides with a particle which is {\sl not} in $S_n$, so that the
number of particles involved doubles at each stage, and the number
in subsystem $S_n$ is $2^n$.

For every collision one of the colliding particles belongs to the
previous subsystem $S_{n-1}$, before the collision, and also to
the subsystem $S_n$ after the collision.  One of the particles is
new, and belongs only to $S_n$.  Starting with particle $P_0$, we
can follow a sequence of collisions and particles leading to a
particle $P_j^{(n)}$ of $S_n$.  For some of these collisions the
particle in this sequence after the collision will be the same as
the particle before the collision.  These collisions are said to
be {\sl direct}.  For others, the particle leaving the collision
will be one of the new ones, and so it will be a different
particle than the one that entered.  These collisions are said to
be {\sl switch} collisions.  In a typical sequence, the number of
direct and switch collisions is roughly equal.

Now consider the gas dynamics.

\subsection{Dynamics of the Arnold gas}

First consider the dynamics of the first collision.  Let ${X}_0$
and ${X}_1$ be the initial state (coordinate and momentum) of
$P_0$ and $P_1$, and let ${X}_0'$ and ${X}_1'$ be the final states
of these particles.

Let $M$ be the matrix of the Arnold cat map \cite{OTT}: \beq
\label{cat} M= \left[
\begin{array}{clcr}
1 & 1\\
1 & 2
\end{array}
\right] \eeq

with eigenvalues \beq \lambda_\pm = {3\pm\sqrt{5}\over 2}. \eeq
Then the equations of the collision, in terms of the centers of
mass and the relative coordinates are \beq \label{kk}
{X}_0'+{X}_1' = {X}_0+{X}_1,\h10 {X}_0'-{X}_1'
 = M({X}_0-{X}_1)
\eeq and in terms of the states of the individual particles are
\beqa { X}_0' &=& {(I+M){X}_0\over 2}+{(I-M){X}_1\over 2} = K_+
{X}_0
+ K_- {X}_1,\\
{X}_1'& =& {(I-M){X}_0\over 2}+{(I+M){X}_1\over 2} = K_- {X}_0 +
K_+ {X}_1 \eeqa

In this collision the linear dependence of the final state of a
particle on its initial state is given by the matrix
$K_+=(I+M)/2$.  This is the {\sl direct} matrix.  The dependence
of the final state of a particle on the initial state of the other
particle is given by the {\sl switch} matrix $K_-=(I-M)/2$.

Using direct and switch matrices we can obtain the linear
dependence of the state of any particle on the initial state $X_0$
of $P_0$.  For a particle $P_j^{(n)}$ of subsystem $S_n$, it has
the form \beq { X}_j^{(n)} = \big(K^+\big)^{n_1}
            \big(K^-\big)^{n_2}{X}_0 + {Y}_0,
\eeq where \beq n_1 + n_2 = n, \eeq Here $n_1$ is the number of
direct matrices and $n_2$ is the number of switch matrices in the
sequence of particles starting with $P_0$ and finishing with
$P_j^{(n)}$. ${Y}_0$ is independent of ${X}_0$ and represents the
initial states of all the other particles of $S_n$.

The value of $n_1$ and thus of $n_2$ depends on the particle
$P_j^{(n)}$.  If it is the same particle as $P_0$, then there are
no switches and $n_1=n$, $n_2=0$.  If the sequence of particles is
a new particle at every stage, from the beginning to the end, then
there are $n$ switches and $n_1=0$, $n_2=n$.  The others lie
between these two extremes.  The number of times a pair
$(n_1,n_2)$ occurs is given by the number of switches, and this
forms a binomial distribution, so the mean values are given by
\beq n_1/n\approx 1/2 \approx n_2/n \hx{(mean values)}, \eeq and
the deviation from this mean becomes relatively small as the
number of collision $n$ increases.

\subsection{Bounds on dilation factors}
Because every collision is represented by a linear map, the same
linear relations hold for displacements $\Delta {X}$ in ${X}$ as
for ${X}$ itself, except for additive constants like ${Y}_0$. So
if the initial state of $P_0$ is displaced by $\Delta {X}_0$, then
the corresponding displacement in $\Delta {X}_j^{(n)}$ is given by
\beq \Delta {X}_j^{(n)} = \big(K^+)^{n_1}\big(K^-\big)^{n_2}\Delta
{X}_0, \eeq where it is assumed that the initial state of every
other particle is held constant.

For a single operation of the Arnold cat map $M$, the eigenvalues
 $\lambda_\pm$ and corresponding normalized eigenvectors
${\bf \xi}_\pm$ are given by \beq M{\bf \xi}_\pm = \lambda_\pm{\bf
\xi}_\pm,\h10 \lambda_\pm = \half(3\pm\sqrt{5}). \eeq

The dilation of the displacement is given by \beq \left|{\Delta
{X}_j^{(n)}\over\Delta {X}_0}\right|, \eeq and this depends on the
direction of $\Delta {X}_0$.  For simplicity, suppose it is in the
direction of ${\bf \xi}_+$, so that \beq \Delta {X}_0 =
\epsilon{\bf \xi}_+, \eeq where $\epsilon$ is the amplitude of the
initial displacement. Now ${\bf \xi}_+$ is an eigenvector of $K^+$
and of $K^-$ as defined in equation (\ref{kk}), \beqa
K^+{\bf \xi}_+ &=& k^+{\bf \xi}_+\\
K^-{\bf \xi}_+ &=& k^-{\bf \xi}_+. \eeqa and the corresponding
eigenvalues are given by \beqa
k^+&=&{1+\lambda_+\over 2}= {5+\sqrt{5}\over 4}\\
k^-&=& {1-\lambda_+\over 2}=-{1+\sqrt{5}\over 4} \eeqa

We also need \beq |k^+ k^-| = 1+{3\over 8}(\sqrt{5}-1) \approx
1.46. \eeq The approximate mean dilation for the displacement of a
single particle after $n$ collisions is therefore \beq
 \left|{\Delta {X}_j^{(n)}\over\Delta {X}_0}\right|  \approx |k^+
k^-|^{n/2} \approx 1.46^{n/2}\approx 1.2^n >1 \eeq The important
thing to notice here is that the mean displacement for the state
of any particle of a subsystem at any later time is greater than
the original displacement for the state of $P_0$.

Now consider these changes as changes in the state of the whole
gas. The initial displacement has magnitude $\epsilon$.  The final
displacement in the phase space of the entire gas has a magnitude
equal to the square root of the sum of the squares of the
displacements of each particle.  The number of particles of $S_n$
is $2^n$, so the dilation for the whole gas is bounded below by
the inequality \beq \hbox{dilation for gas}>\sqrt{2^n}= 2^{n\over
2}. \eeq The magnitude of the displacement in the phase space of
the whole gas of $N$ particles more than doubles in every two
stages, and becomes significant after fewer than ${}{2}\log_2 N$
collisions, in the sense described in the next section.

Now we come to the question of times.  The stages correspond to
different times for different collisions, but the time for $n$
collisions is roughly the same when $n$ is sufficiently large, and
approximately equal to $n\Delta t$ where $\Delta t$ is the mean
time between two collisions of a single particle.

The time $T_s$ for the displacement to become significant, in the
sense that a typical particle of the gas of $N$ particles at time
$t=T$ has roughly the same displacement as $P_0$ has at time $t=0$
is then \beq T_s\approx \Delta t\log_2 N \eeq
\section{Fluctuation of the density}

This section is devoted to discuss the fluctuation of the Arnold
gas density in  phase space. An exponent for the Fourier
components of the density is defined.  This exponent gives a means
of characterizing the fluctuation of the density.

The phase density for a system comprised of $N$ particles in a
unit square is given by \beq n({X},t)=\sum_{i=1}^N
\delta({X}(t)-{X}_i(t)) \eeq where

\beq \label{cat1} X(t)= \left[
\begin{array}{clcr}
x_i(t) \\
p_i(t)
\end{array}
\right] \eeq here $x_i(t)$ and $p_i(t)$ are the position and
momentum of $i$th particle, respectively. The use of the notation
$n(X,t)$ for the density should not be confused with the number of
iterations. It is easy to see that \beq \int
n({X},t)d{X}=\sum^N_{i=1}\int \delta({X}-{X}_i)d{X}=N \eeq where
$d{X}$ denotes $dxdp$. The Fourier expansion of $n({X},t)$ is
given by: \beq n({X},t)=\sum_{i=1}^N \delta({X}(t)-{X}_i(t))=
L^{-2}\sum_{k} n_{k}(t) \exp(i {k}\cdot{X}(t)) \eeq with \beq
n_{k}(t)=\int n({X},t) \exp\left[-i{k}\cdot {X}(t)\right] d{X}
=\sum_{i=1}^N\exp[-i{k}\cdot {X}_i(t)] \eeq where $L$ (We choose
$L=1$) is the length of the square containing $N$ particles and
$\sum_{k}$ stands for the sum over all possible discrete values of
${k}$ allowed by the imposed boundary condition.

Now we consider the probability density $\ti{n}({X},t)$: \beq \ti
n({X},t)=\frac{1}{N}n({X},t) \eeq

We are now in the position to compute a  bound on the ratio of
the Fourier component $\ti n_{k}(t)$ and  the initial displacement
of a particle, say, particle $P_0$.  To do so, first, note that
\beq \ti n_{k}(t)=\frac{1}{N}n_{k}(t) \eeq Hence, $\Delta\ti
n_{k}(t)=\frac{1}{N}\Delta n_{k}(t)$. Note that
 $t$ is an integer representing the number of iterations. So
the ratio is given by \beq \left|\frac{\Delta \ti n_{k}(t)}{\Delta
{X}_0}\right| =\left|\frac{\sum_ie^{i{k}\cdot {X}_i^{(t)}}{k}\cdot
\Delta {X}_i^{(t)}}{N\Delta {X}_0}\right| \eeq Note that \beq
\label{fac} {k}\cdot\Delta {X}_i^{(t)} =
 {k}\cdot\big(K^+)^{n_1}\big(K^-\big)^{n_2}\Delta {X}_0.
\eeq

If we consider the displacement of the particle $P_0$ in the
direction of $\xi_+$: $\Delta {X}_0=\epsilon \xi_+ $, then the
equation (\ref{fac}) becomes \beq {k}\cdot\Delta
{X}_j^{(t)}\approx (1.2)^t\epsilon {k}\cdot \xi_+ \eeq

An exponent for the Fourier component ${\ti n}_{k}(t)$ can be
defined as \beq \label{lya}
\lambda=\ln\left|\frac{\sum_ie^{i{k}\cdot
{X}_i^{(t)}}({k}\cdot\xi_+)}{N}\right|^{1\over t}+\ln(1.2). \eeq
Note that the exponent $\lambda$ plays the similar role to the
Lyapunov exponent. But unlike the Lyapunov exponent, $\lambda$ is
not always positive.  For sufficiently large $t$, the second term
in Eq. (\ref{lya}) is dominant.  So we have \beq \label{exp}
\lambda \approx \ln(1.2) \approx 0.18> 0 \eeq

Hence, in the long time limit, with (\ref{lya}) and (\ref{exp}),
the dilation of the Fourier component $\ti n_{k}(t)$ can be
written as \beq \Delta \ti n_{k}(t)\approx \epsilon e^{\lambda t}.
\eeq

It should be emphasized that the exponent $\lambda$  may not be
positive at the early stage as can be seen from (\ref{lya}). It
only becomes positive when the collisions have significantly
influenced the whole gas. This is consistent with our expectation.

It is interesting to make some rough estimations of the time
scale. We assume that gas at room temperature($T\approx 300 K$)
and atmosphere pressure ($p\approx 10^5 {\rm Nm}^{-2}$) is contained
in a square with area $100^{-2}{\rm m}^2$, that is, $L=1{\rm cm}$.
The number of Arnold gas
molecules is about $2.5\times 10^{19}$.  The mean free path is
$l_m = 2\times 10^{-7}{\rm m}$. The mean speed of molecules is
$v_{m}=4\times 10^2$m/sec. Then the mean free time is $ t_m
=l_m/v_m \approx 5\times 10^{-10}\ {\rm sec}$ .  Now we see that
in one second, there are approximately $2\times 10^9$(iterations)
collisions. We see that the small changes in the phase space of a
single particle can produce the exponential difference in the
particle probability density $\Delta \ti n_{k}(t)\approx \epsilon
e^{\lambda t}$.

Finally, let's take a look at the familiar example of light
scattering.  For a volume element $\Delta V$ which is of the
dimension of the order of the wavelength of visible light($\approx
5\time 10^{-7}$m), the fluctuation are significant, as shown
above. The Rayleigh scattering is due to the fluctuation of the
particle density. This is in turn responsible for the blue of the
sky.

\section{Concluding comments}

We have constructed a classical model of an equilibrium gas to
represent the classical stage of a generalized quantum measurement. An
exponent is used to characterize the fluctuation of the gas
density relative to the initial displacement of a single particle.
To be specific, we have shown that density of the Arnold gas is
highly sensitive to a disturbance of the initial position and
momentum of one particle.

In some sense, the model looks artificial, because of the
following differences between the model gas and a real gas of
molecules:

(i)  Normal gases are 3-dimensional, not 1-dimensional.

(ii) An ordinary collision between two molecules of a gas does not
resemble any kind of linear cat map, even a 6-dimensional cat map.
A collision of the cat map here corresponds to a collision and
subsequent drift in a real gas.

(iii) After a sufficient number of collisions, the number of molecules
in the system $S_n$ affected by a displacement of $P_0$ does not
double at every two stages, because molecules of a subsystem can
collide with each other.  The number of molecules in $S_{n+1}$ is then
less than $2^{n+1}$.  Some of the particles of the subsystem are
affected as a result of two or more sequences of collisions, between
different particles.  For a real gas like the atmosphere, ignoring the
effects of radiation, the particles in the subsystem affected by a
displacement $\Delta {X}_0$ is determined by the speed of sound, and
increases asymptotically as the cube of the time.  Because the number
of particles in the real gas is less than for the model gas, the
dilation factor is larger for the real gas.

If the displacement in the phase space of the two particles after
a collision is more than (not necessarily more than double) the
displacement in the phase space of one of the particles before the
collision, when the displacement of the other particle is zero,
then the displacement of the phase point of the gas grows
exponentially at each stage.  A typical ratio is more difficult to
work out for the nonlinear dynamics of real collisions, partly
because `collision' is not clearly defined for potentials of
infinite range.

For a particle which receives a displacement as a result of two
different sequences of collisions, it may be a good approximation
to assume that these displacements are statistically independent,
in which case the resultant displacement is equivalent to
displacements of different molecules.

The details of these considerations go beyond the scope of this
paper.  The present paper only serves as a first step towards the
generalized quantum measurement theory of equilibrium systems. Of
course, the present paper is not complete because we have ignored
the relation between quantum fluctuations and classical
fluctuations at the ambiguous boundary between the `classical' and
the `quantum' domains.

There are many situations in which a generalized quantum measurement
is of interest.  A remarkable example of this situation arises in
the early universe context in which the density fluctuation is
important for the early evolution of the universe. Crudely
speaking, the long wavelength radiation could serve as the
environment field whereas the short wave-length as quantum modes
\cite{LIN, HBL,HBL1}. The interaction between those different
modes will be important for the development of early universe such
as vacuum particle creation and structural formation.

One of John Bell's major objections to `quantum measurement' can
be overcome by generalizing the definition to include processes in
the big world.  With this definition, quantum measurement takes
place in those equilibrium systems for which the classical motion
is chaotic, even though the measurement cannot be used in that
case to get detailed information about individual quantum states.
Consequently the dynamics of quantum measurement has universal
significance and so have its properties.

This paper is a first step towards to a 
theory of generalized quantum measurement for equilibrium systems.

\section*{Acknowledgments}

The authors acknowledge helpful and stimulating communications
with N. Gisin.  This work was supported by the Leverhulme
foundation and the ESF.


\end{document}